\documentclass[twocolumn,superscriptaddress, amsmath, amssymb,  prd, showpacs, floatfix ,preprintnumbers, natbib]{revtex4-2}

\usepackage{graphicx}
\usepackage{here}
\usepackage{dcolumn}
\usepackage[usenames]{color}
\usepackage[colorlinks=true,linkcolor=blue,citecolor=blue,urlcolor=blue]{hyperref}
\usepackage{textfit}
\usepackage{bbm}
\usepackage{natbib}
\usepackage{caption}
\usepackage{subcaption}
\usepackage{mathtools}

\usepackage{dcolumn}
\usepackage{bm}
\usepackage{epsfig}
\usepackage{soul,xcolor}
\usepackage[normalem]{ulem}
\usepackage[makeroom]{cancel}

\begin{document}
\preprint{ADP-20-34/T1144}
\title{Censored EM algorithm for Weibull mixtures: application to arrival times of market orders }

\author{Markus Kreer}
\affiliation{CSSM, Faculty of Sciences, School of Physical Sciences, Department of Physics, University of Adelaide, 5005,
Adelaide, Australia.} 
\affiliation{CAMPUSERVICE GmbH, Servicegesellschaft der Johann Wolfgang Goethe-Universitat Frankfurt, Rossertstrasse 2, 
60323 Frankfurt am Main, Germany}
\affiliation{Feldbergschule, Oberhochstadter Str., 20, 61440, Oberrursel (Taunus), Germany}

\author{Ay{\c s}e K{\i}z{\i}lers\"u}
\affiliation{CSSM, Faculty of Sciences, School of Physical Sciences, Department of Physics, University of Adelaide, 5005,
Adelaide, Australia.} 
\email{ayse.kizilersu@adelaide.edu.au}

\author{Anthony W. Thomas}
\affiliation{CSSM, Faculty of Sciences, School of Physical Sciences, Department of Physics, University of Adelaide, 5005,
Adelaide, Australia.}

\begin{abstract}
In a previous analysis the problem of ``zero-inflated'' time data (caused by high frequency trading in the electronic order book) was handled by left-truncating the inter-arrival times. 
We demonstrated, using rigorous  statistical methods, that the Weibull distribution describes the corresponding stochastic dynamics for all inter-arrival time differences except in the region near zero. 
However, since the truncated Weibull distribution was not able to describe the huge ``zero-inflated'' probability mass in the neighbourhood of zero (making up approximately 50\% of the data for limit orders), it became clear that the entire probability distribution is a mixture distribution of which the Weibull distribution is a significant part. Here we use a censored EM algorithm to analyse data for the difference of the arrival times of market orders, which usually have a much lower percentage of zero inflation, for four selected stocks trading on the London Stock Exchange.
\end{abstract}

\maketitle

\section{Introduction}
Electronic Order Book (EOB) trading is now common to most stock exchanges. A set of EOB data from the FTSE for the period June to September 2010 was described and partly analysed in Ref.\cite{Kizilersu2016}. In this data set timestamps of changes in the EOB were given in milliseconds and because of the huge volume of EOB trading by (ultra-)high frequency trading algorithms, taking place on a microsecond scale, differences of timestamps appeared to be zero. This is why the timestamp time series appeared to be  ``zero-inflated''. Because of this rounding error, all differences in timestamps, $\Delta t$, are mapped to our actual data set as follows
\begin{eqnarray}
\Delta t \in (0.0,0.5) =\mathcal{I}_{1} &\longrightarrow& 0 = c_1
\nonumber \\
\Delta t \in [0.5,1.5) =\mathcal{I}_{2} &\longrightarrow& 1 = c_2
\nonumber \\
\Delta t \in [1.5,2.5) =\mathcal{I}_{3} &\longrightarrow& 2 = c_3
\nonumber \\
& etc. &
\nonumber
\end{eqnarray}
where the integer numbers $c_1, c_2,...$ are the rounded time stamp differences to the precision of milliseconds.

In Ref.\cite{Kizilersu2016} it was demonstrated that sets of non-negative time differences between subsequent market orders (MO) with $\Delta t >10 $ milliseconds describe a random variable $X$ that could be fitted to parametric distributions, such as a left-truncated Weibull distribution, which passed a rigorous set of goodness-of-fit tests, such as Kolmogorof-Simirnov, Anderson-Darling, Cramer von Misses and Kuiper tests\footnote{The analysis in Ref.\cite{Kizilersu2016} also included limit orders (LO) but this will not be the topic of our letter.}.


Here we analyse the entire set of observations of the random variable $X$, including the "zero-inflated" ones, by starting from a Weibull distribution for the whole range
\begin{eqnarray}
f_i(x|\alpha_i,\beta_i) = \frac{\beta_i}{\alpha_i} \left(\frac{x}{\alpha_i}  \right)^{\beta_i-1} e^{-\left(\frac{x}{\alpha_i}  \right)^{\beta_i}} \, ,
\label{eq:Weibull_pdf}
\end{eqnarray}
where $\alpha_i>0$ is the scale parameter and $\beta_i>0$ the shape parameter. Note that for $\beta_i=1$ we recover the exponential distribution. We denote the parameter vector by $\theta_i=(\alpha_i,\beta_i)$ and use the subscript index $i$ to denote various Weibull or exponential distributions with different parameter vectors $\theta_i$.

Now, the attempt to analyse the entire data set including the ``zero-inflated'' part (which sometimes was even more than half of the observed data set) requires a different procedure. One promising approach is to apply interval censoring to the observed data for the intervals of small time differences $\Delta t$: we use the above intervals $\mathcal{I}_{1}, \mathcal{I}_{2}, \mathcal{I}_{3}, ...$ together with the information of how many observations belong to them, i.e. $N_1, N_2, N_3,...$. In this notation we can write an observed sample of a positive random variable $X$ as
\begin{eqnarray}
x_1, &x_2, \cdots&,x_n,x_{n+1},x_{n+2},..., x_N \nonumber\\
 &=& x_1, x_2,...,x_n, \underbrace{c_1,c_1,...,c_1}_{N_1 \text{times}},...,
\underbrace{c_L,c_L,...,c_L}_{N_L \text{times}}  \, ,
\label{eq:datasample}
\end{eqnarray}
where, after grouping, all observations  with index bigger than $n$ have been censored (with $L$ censoring intervals).

Despite censoring, the random variable $X$ itself will be assumed to come from a mixed distribution with a  mixture of $M$ components. As a consequence of the findings in Ref.\cite{Kizilersu2016}, we expect that a significant component of this mixed distribution should come from a Weibull distribution. 
Similar results where found in Ref.\cite{Scalas2008}: the authors found that the differences between subsequent MO's for 30 DJIA stocks from the NYSE in October 1999 can be well described by Weibull distributions.
Thus we want to investigate here mixtures of the following kind
\begin{eqnarray}
f(x|\theta) =  \sum_{i=1}^M \Pi_i f_i(x|\theta_i)
\label{eq:pdf_mix}
\end{eqnarray}
where the weights, $\Pi_i$, satisfy $\Pi_i\geq 0$ with $\sum_{i=1}^M \Pi_i = 1 $ and the probability density functions $f_i(\cdot|\theta_i)$ are given by Eq.(\ref{eq:Weibull_pdf}), where the parameters $\theta_i$  need to be estimated by maximum likelihood estimation or other methods\footnote{In the following the index "i" stands for the mixture components and the index "j" for the number of observation.}. Defining $\Theta=(\theta_1,...,\theta_M)$, the log-likelihood function $\mathcal{L}$ for $L$ censoring intervals $\mathcal{I}_{1},...,\mathcal{I}_{L}$ is now given as in Ref.\cite{Kendall1979} by
\begin{widetext}
\begin{eqnarray}
\mathcal{L} = f(x_1|\Theta)\cdots\ f(x_n|\Theta)\cdot \left(\int_{\mathcal{I}_1} ~dy f(y|\Theta)\right)^{N_1} \cdots \left(\int_{\mathcal{I}_L} ~dy f(y|\Theta) \right)^{N_L}
\label{eq:likelihood}
\end{eqnarray}
\end{widetext}
which is usually a difficult expression to handle. However, given our data sample Eq.(\ref{eq:datasample}), the estimation problem would be a standard maximum likelihood estimation (MLE) problem if we knew by some indicator $z_{ij}$ (taking values 0 or 1) denoting which observation $x_j$ belongs to which probability density function $f_i(\cdot|\theta_i)$ and likewise some indicator $\tilde{z}_{i\ell}$ (also taking values 0 or 1) which censored observation $c_{\ell}$ belongs to which distribution. In this case, we would just group the observations and thus factorize the likelihood $\mathcal{L}=\mathcal{L}_1\cdots \mathcal{L}_M$ and separately maximize each group likelihood $\mathcal{L}_i$ corresponding to one mixture component $f_i(\cdot|\theta_i)$. 

The expectation-maximization (EM) algorithm Ref.\cite{McLachlan, Dempster1977, Redner1984} will be used to iteratively generate estimates for the indicators $z_{ij}$, for uncensored observations, $x_j$ and $\tilde{z}_{i\ell}$ for censored observations,  $c_{\ell}$, and solve the estimation problem (provided it converges). The problem for censored mixtures has been discussed in some detail in Ref.\cite{Chauveau1995}. In our study we apply this analysis for mixtures of exponential and Weibull distributions and study market order (MO) inter-arrival times. In section 2 we give a very brief review of the censored EM algorithm as given by Ref.\cite{Chauveau1995} and in section 3 the relevant equations for Weibull distributions are given. Section 4 summarizes our results and conclusions.

\section{Censored EM algorithm for mixtures in a nutshell}
The EM algorithm is an iterative procedure where an expectation-step (E-step) tries to estimate the unobservable indicators $z_{ij}$ and $\tilde{z}_{i\ell}$, respectively, and then uses the result in the maximization step (M-step) to estimate the parameters of the mixtures by an MLE. After the M-step there is  another E-step and so on. Here we follow closely Ref.\cite{Chauveau1995}, where a hint is given that this iteration converges for certain function families  (including exponential and Weibull distribution functions) in a certain limited parameter range. In the following sections the integer index $k$ denotes the number of the iteration.

\subsection{The E-step}
Using the above notation and assuming that the parameter vector $\theta_i^{(k-1)}$ for each mixture component $i$ are known from a previous step,  Eq.(3.3)  in the Ref.\cite{Chauveau1995} provides   the following term for the uncensored observations, $j=1,...,n$ for the mixture component $i$
\begin{eqnarray}
z_{ij}^{(k)} 
& = & \frac{f_i(x_j|\theta_i^{(k-1)})}{\sum_{i=1}^M \Pi_i^{(k-1)} f_i(x_j|\theta_i^{(k-1)})} \cdot \Pi_i^{(k-1)}
\label{eq:z_ij}
\end{eqnarray}
and for the censored observations in the censoring intervals $\mathcal{I}_{\ell}=[\xi_{\ell-1},\xi_{\ell})$ with $\ell=1,2,...,L$, Eq.(3.4) in the Ref.\cite{Chauveau1995} provides the following expression for the mixture $i$th component
\begin{eqnarray}
\tilde{z}_{i\ell}^{(k)} 
& = & \frac{\int_{\mathcal{I}_{\ell}} dy~ f_i(y|\theta_i^{(k-1)})}{\sum_{i=1}^M \Pi_i^{(k-1)}\int_{\mathcal{I}_{\ell}} dy~  f_i(y|\theta_i^{(k-1)})} \cdot \Pi_i^{(k-1)}
\label{eq:Z_il}
\end{eqnarray}

\subsection{The M-step}
The sample size, $N=n+\sum_{\ell=1}^L N_{\ell}$, is the number of all observations (uncensored and censored) and the new weights after  the E-step are given by the following update-rule (Eq.(3.7) of Ref.\cite{Chauveau1995}):
\begin{eqnarray}
\Pi_i^{(k)} & = & \hspace{0.3cm} \frac{1}{N} \sum_{j=1}^n \frac{f_i(x_j|\theta_i^{(k-1)})}{\sum_{i=1}^M \Pi_i^{(k-1)} f_i(x_j|\theta_i^{(k-1)})} \cdot \Pi_i^{(k-1)} 
\nonumber
\\
&& + \frac{1}{N}\sum_{\ell=1}^L N_{\ell}\frac{\int_{\mathcal{I}_{\ell}} dy~ f_i(y|\theta_i^{(k-1)})}{\sum_{i=1}^M \Pi_i^{(k-1)}\int_{\mathcal{I}_{\ell}} dy~  f_i(y|\theta_i^{(k-1)})} \cdot \Pi_i^{(k-1)}
\, .
\nonumber
\end{eqnarray}
Note that this update-rule is just same as the following formula
\begin{eqnarray}
\Pi_i^{(k)} = \frac{1}{N} \left( \sum_{j=1}^n z_{ij}^{(k)}+ \sum_{\ell=1}^L N_{\ell} \tilde{z}_{i\ell}^{(k)}\right)
\label{eq:Pi_k}
\end{eqnarray}

Now, 
the following expression needs to be maximised with respect to the parameter vectors $\theta_i^{(k)}$, where the index $i$ refers to the mixture and $k$ to the actual number in the iterative proceedure
\begin{widetext}
\begin{eqnarray}
\sum_{i=1}^M \sum_{j=1}^{n} z_{ij}^{(k)} \log{f_{i}(x_j|\theta_{i}^{(k)}) } + \sum_{i=1}^M \sum_{\ell=1}^L N_{\ell}~ \tilde{z}_{i\ell}^{(k)} \int_{\mathcal{I}_{\ell}} dy ~\log{f_{i}(y|\theta_{i}^{(k)})} h_i(y|c_{\ell},\theta_i^{(k-1)})
\nonumber \\ 
\label{eq:Q_term}
\end{eqnarray}
\end{widetext}
Here, the conditional density function on the censoring interval $\mathcal{I}_{\ell}$ is given by Ref.\cite{Chauveau1995} Eq.(3.5) 
\begin{eqnarray}
 h_i(y|c_\ell,\theta_i^{(k-1)}) = \frac{f_i(y|\theta_i^{(k-1)})}{\int_{\mathcal{I}_{\ell}}dy~  f_i(y|\theta_i^{(k-1)})} 
\end{eqnarray}

The MLE equations will be obtained by differentiating Eq.(\ref{eq:Q_term}) with respect to the components of the parameter vector $\theta_i^{(k)}$. Note that in this expression terms like $\sum_{i,j} z_{ij}^{(k)} \log{\Pi_i^{(k)}}$ and the normalization condition  $\mu \cdot \left( \sum_{i} \Pi_i^{(k)}-1\right)$ are not given because they depend only on the parameter vector $\theta_i^{(k-1)}$ from the previous iteration and are irrelevant for the maximization with respect to the parameter vector $\theta_i^{(k)}$. Note also that the maximization problem decouples into  independent maximization problems for each individual probability distribution.


\section{Implementation of the  censored Weibull mixtures}
%
For simplicity we now consider a 2-component mixture consisting of an exponential distribution, denoted by $i=1$, and a general Weibull distribution, denoted by $i=2$, as given in Eq.(\ref{eq:Weibull_pdf}). Note that here the $\alpha$ and $\beta$ have a subscript $i$ for the mixture and a superscript $(k)$ for the iteration step in the EM algorithm. The expression corresponding to Eq.(\ref{eq:Q_term}) is given in Appendix A. Our algorithmic results extend the results as given by Ref.\cite{Jewell1982} for exponential mixtures. It is clear how to modify our computations for arbitrary mixtures, e.g. $(p+r)$ mixtures consisting of $p$ exponentials and $r$ Weibulls,with $p,r=0,1,2,...$

\subsection{MLE equations for M-step}
Maximising the expression in Appendix A, Eq.(\ref{eq:MLE}), with respect to the first parameter, $\alpha_1^{(k)}$ leads for the exponential distribution (index $i=1$) to an explicit solution 
\begin{eqnarray}
\alpha_1^{(k)} = \frac{\sum_{j=1}^n z_{1j}^{(k)} x_j + \sum_{\ell=1}^L N_{\ell} \tilde{z}_{1\ell}^{(k)} C_{1\ell}^{(k-1)}}{\sum_{j=1}^n z_{1j}^{(k)}+\sum_{\ell=1}^L N_{\ell} \tilde{z}_{1\ell}^{(k)}}
\label{eq:exponential_MLE}
\end{eqnarray}
where the quantity $C_{1\ell}^{(k-1)}$ is defined in Appendix A, Eq.(\ref{eq:C_ell}).
  
The MLE equations  for the Weibull distribution (index $i=2$) are obtained by computing  $\frac{\partial}{\partial\alpha_2^{(k)}}$ and equating the expression to 0,
\begin{widetext}
\begin{eqnarray}
0 & = & \sum_{j=1}^{n} z_{2j}^{(k)}  \left[ -1 + \left(\frac{x_j}{\alpha_2^{(k)}}\right)^{\beta_2^{(k)}} \right]
\nonumber
\\
&+& \sum_{\ell=1}^L N_{\ell}~ \tilde{z}_{2\ell}^{(k)}  \left\{ -1+\left(\frac{\alpha_2^{(k-1)}}{\alpha_2^{(k)}}\right)^{\beta_2^{(k)}} 
\frac{ \Gamma\left(\frac{\beta_2^{(k)}}{\beta_2^{(k-1)}}+1 ,\zeta_{\ell-1}\right)-\Gamma\left(\frac{\beta_2^{(k)}}{\beta_2^{(k-1)}}+1 ,\zeta_{\ell}\right) }{e^{-\zeta_{\ell-1}}-e^{-\zeta_{\ell}}} \right\}
\nonumber 
\\
&&
\label{eq:Weibull_alpha}
\end{eqnarray}
\end{widetext}
and likewise for  $\frac{\partial}{\partial\beta_2^{(k)}}$
\begin{eqnarray}
&&0= \sum_{j=1}^{n} z_{2j}^{(k)}  \left[ \frac{1}{\beta_2^{(k)}} + \log{\frac{x_j}{\alpha_2^{(k)}}} - \left(\frac{x_j}{\alpha_2^{(k)}}\right)^{\beta_2^{(k)}} \log{\frac{x_j}{\alpha_2^{(k)}}}\right]
\nonumber
\\
&&+  \sum_{\ell=1}^L N_{\ell}~ \tilde{z}_{2\ell}^{(k)}  \left\{ \frac{1}{\beta_2^{(k)}} + \log{\frac{\alpha_2^{(k-1)}}{\alpha_2^{(k)}}} + \frac{D_{2\ell}^{(k)}}{e^{-\zeta_{\ell-1}}-e^{-\zeta_{\ell}}} \right\}\,\, 
\label{eq:Weibull_beta}
\end{eqnarray}
The auxiliary functions  $D_{2\ell}^{(k)}$, as well as $\zeta_{\ell}$, are defined in the Appendix A. Reference \cite{Elmahdy2013} has obtained similar results for the complete finite mixture of Weibull distributions albeit without the censoring terms displayed here.

\subsection{Algorithmical implementation }
We use one censoring interval only, $\mathcal{I}_{1}=(0,0.5)$ and $L=1$ to handle our zero-inflated data sets (see Ref.\cite{Kizilersu2016}). Also in the spirit of Ref.\cite{Efron1967}, as a practical approximation to simplify the solution of these non-linear equations, we use a self-consistency assumption by setting the ratios $\alpha_2^{(k-1)}/\alpha_2^{(k)}$ and $\beta_2^{(k-1)}/\beta_2^{(k)}$ equal to 1 in the MLE equations for the Weibull parameters. This considerably simplifies the MLE equations, 
Eqs.~(\ref{eq:Weibull_alpha})--(\ref{eq:Weibull_beta}), and in our experience leads to rapid convergence in most cases.\footnote{A comparison to the direct maximisation of the terms in Eq.(\ref{eq:Q_term}) as given in the Appendix A leads to comparable results and might justify our trick to speed-up convergence. Note that if we already knew the exact solution $\alpha_2^{\infty}$ and $\beta_2^{\infty}$, i.e. if we started with the true fixed point, these equations would be trivially satisfied.}
For the EM-algorithm we proceed as follows\footnote{We describe the censored EM-algorithm for a mixture of 1 exponential + 1 Weibull. The generalisation to arbitrary mixtures is obvious}:
\begin{enumerate}
\item At $k=0$ initialize mixture weights $\Pi_i^{(0)}$ and initial values $\alpha_1^{(0)}, \alpha_2^{(0)}, \beta_2^{(0)}$
\item Compute $z_{ij}^{(k)},\tilde{z}_{i\ell}^{(k)}, \Pi_i^{(k)}$ using Eq.(\ref{eq:z_ij}), (\ref{eq:Z_il}), (\ref{eq:Pi_k}) for $k=1,2,...$
\item Compute $\alpha_1^{(k)}$ using Eq.(\ref{eq:exponential_MLE}) for exponential distribution for $k=1,2,...$
\item Compute $\alpha_2^{(k)}$ using Eq.(\ref{eq:Weibull_alpha})  for Weibull distribution putting here only $\beta_2^{(k)}=\beta_2^{(k-1)}$ for $k=1,2,...$
\item Compute $\beta_2^{(k)}$ using Eq.(\ref{eq:Weibull_beta}) for Weibull distribution for $k=1,2,...$
\item Compute the current log-likelihood from Eq.(\ref{eq:pdf_mix})--(\ref{eq:likelihood})
\item If the absolute value of ``log-likelihood at step $k$ minus log-likelihood at step $k-1$'' is bigger than $\varepsilon>0$, then put $k\rightarrow k+1$ and go back to step 2., otherwise terminate.
\end{enumerate}

In our experience this version of the censored EM algorithm (based on the MLE equations) converges  sufficiently fast to a desired maximum solution for suitable initial conditions. We have taken $\varepsilon=10^{-5}$ and start with equal mixtures setting $\beta=1$ and take values of $\alpha$ motivated by our previous study Ref.\cite{Kizilersu2016}. When testing the algorithm, its results have been cross-checked by the corresponding algorithm which uses a direct maximisation of the objective function given in Eq.(\ref{eq:Q_term}) in the M-step. Usually the results obtained both way agree fairly well.  In Ref.\cite{Chauveau1995} it is claimed that there is local convergence almost surely.

\section{Analysis of MO arrival times and model selection}
\subsection{A first approach: ``naive'' analysis of entire data set}
Here we take all time stamps of a given stock from 1st June to 30th September 2010 into one large sample and fit various mixture models to the arrival times (i.e. the difference of subsequent timestamps) using the algorithm described above. This analysis would be sensible if the stochastic process were stationary and the sample data were independent identically distributed (i.i.d).
In Table~\ref{tab:ave_logL} we provide the log-likelihood per data point to obtain a first idea about the best model. The percentage numbers in round brackets denote the proportion of those data points which have been censored. Note that the quantity ``log-likelihood per data point'', or ``average log-likelihood'', corresponds to a negative Shannon entropy per event because for large sample size $N$ and i.i.d. data we have under the usual consistency property of the maximum likelihood estimator
\begin{eqnarray}
\frac{1}{N}\log{\mathcal{L}} \sim \mathbb{E}(\log{p})=\int \log{p}\cdot dp
\nonumber
\end{eqnarray}
The physical meaning of this quantity is the ``information content'' or ``surprisal'' when a new MO enters the EOB.
\begin{table}[htbp]
  \centering
  \caption{Average log-likelihood}
  \begin{ruledtabular}
    \begin{tabular}{lrrrrr}
          &       & \multicolumn{1}{l}{RIO} & \multicolumn{1}{l}{BARC} & \multicolumn{1}{l}{RRLN} & \multicolumn{1}{l}{ABFLN} \\
          &       & (2.5\%) & (2.3\%) & (2.5\%) & (2.1\%) \\
   Model & \multicolumn{1}{l}{dof} &       &       &       &  \\ \hline
    1 exp + 1 wbl & 4     & -8.357 & -8.294 & -9.510 & -10.074 \\
    0 exp + 2 wbl & 5     & -8.349 & -8.288 & -9.492 & -10.060 \\
    3 exp + 0 wbl & 5     & -8.422 & -8.394 & -9.792 & -10.323 \\
    2 exp + 1 wbl & 6     & -8.350 & -8.290 & -9.494 & -10.061 \\
    1 exp + 2 wbl & 7     & -8.348 & -8.288 & -9.489 & -10.057 \\
    4 exp + 0 wbl & 7     & -8.360 & -8.300 & -9.511 & -10.081 \\
    0 exp + 3 wbl & 8     & -8.348 & -8.288 & -9.489 & -10.054\\
    3 exp + 1 wbl & 8     & -8.349 & -8.289 & -9.491 & -10.055 \\ 
    5 exp + 0 wbl & 9     & -8.351 & -8.291 & -9.497 & -10.063 \\
    \end{tabular}%
    \end{ruledtabular}
  \label{tab:ave_logL}%
\end{table}%

We see from Table~\ref{tab:ave_logL} that all models seem to yield very similar Shannon entropy measures for an individual stock. Also, depending on the trading activity, the entropies differ: for a heavily traded stock such as RIO or BARC the ``surprise'' is lower than for less actively stocks such as RRLN or ABFLN. A model selection criterion here would be the model with least entropy and thus  a mixture of 3 Weibull distributions seems to be the best choice. 

However, we know from Ref.\cite{Kizilersu2016} that the data are only i.i.d. for smaller subsamples and thus the scale parameters in the exponential and Weibull distribution will exhibit a time dependence. When looking at smaller samples it became customary for model selection rather looking at the above Shannon entropy  to put the sample size in relation to the degrees of freedom (dof) (see e.g. Ref.\cite{Burnham2002}). We have decided to use  the Bayesian Information Criterion (BIC) (Ref.\cite{Kass1995}, Ref.\cite{Burnham2002})
\begin{eqnarray}
\text{BIC} = -2\log{\mathcal{L}}+d\log{N}
\label{eq:BIC}
\end{eqnarray} 
where $\mathcal{L}$ are likelihoods and $d$ is the number of degrees of freedom of the individual model.  From Table~\ref{tab:ave_logL} we can generate hypothetically Table~\ref{tab:ave_BIC} by using Eq.(\ref{eq:BIC}) with $N=200$ for ``larger stocks'', with higher trading activity (RIO, BARC), and $N=100$ for ``smaller stocks'', with lower trading activity (RRLN, ABFLN). These values for $N$ correspond to time intervals of approximately 10 minutes and we have seen in Ref.\cite{Kizilersu2016} that these time intervals yield samples for which the asumption of stationarity of the data set can be somehow justified. We clearly see that now mixtures with low dof are favored, in particular the mixture ``1 exponential + 1 Weibull''. Note also that the suggestion of Ref.\cite{Scalas2007} to model the arrival time distribution as a suitable mixture of exponential waiting times, will be excluded in the model selection using BIC by a too high value for the dof.
\begin{table}[htbp]
  \centering
  \caption{Expected BIC from Table1 with different sample size $N$}
  \begin{ruledtabular}
    \begin{tabular}{lrrrrr}
     &       & \multicolumn{1}{l}{RIO} & \multicolumn{1}{l}{BARC} & \multicolumn{1}{l}{RRLN} & \multicolumn{1}{l}{ABFLN} \\
         &       & (2.5\%) & (2.3\%) & (2.5\%) & (2.1\%) \\
    Model & \multicolumn{1}{l}{dof} &  $N=200$     &  $N=200$     &  $N=100$     & $N=100$ \\ \hline
    1 exp + 1 wbl & 4     & 3363.99 & 3338.79 & 1923.19 & 2035.99 \\
    0 exp + 2 wbl & 5     & 3366.09 & 3341.69 & 1924.89 & 2038.49 \\
    3 exp + 0 wbl & 5     & 3395.29 & 3384.09 & 1984.89 & 2091.09 \\
    2 exp + 1 wbl & 6     & 3371.79 & 3347.79 & 1930.59 & 2043.99 \\
    1 exp + 2 wbl & 7     & 3376.29 & 3352.29 & 1934.89 & 2048.49 \\
    4 exp + 0 wbl & 7     & 3381.09 & 3357.09 & 1939.29 & 2053.29 \\
    0 exp + 3 wbl & 8     & 3381.59 & 3357.59 & 1940.19 & 2053.19 \\
    3 exp + 1 wbl & 8     & 3381.99 & 3357.99 & 1940.59 & 2053.39 \\
    5 exp + 0 wbl & 9     & 3388.08 & 3364.08 & 1947.08 & 2060.28 \\
    \end{tabular}%
    \end{ruledtabular}
  \label{tab:ave_BIC}%
\end{table}%

\subsection{A second approach: Analysis of stationary subsamples}
The previous analysis was naive as we assumed the entire data sample would consist of i.i.d. random variables. This is clearly not the case. As already noticed in Ref.\cite{Kizilersu2016} the volume of trading changes a great deal during a trading day, so that the scale parameter $\alpha$ must also vary. However, in Ref.\cite{Kizilersu2016} Kizilers\"u {\em et al.} argued that for ``small'' subsamples the assumptions of stationarity and i.i.d. might be expected to be justified. We take as subsample size $N=200$ for ``larger stocks'' with higher trading activity (RIO, BARC) and $N=100$ for ``smaller stocks'' with lower trading activity (RRLN, ABFLN).
Motivated by Table~\ref{tab:ave_BIC}, our candidates for possible models are the following mixtures
\begin{itemize}
\item 1 exp + 1 wbl  (dof=4)
\item 0 exp + 2 wbl  (dof=5)
\item 3 exp + 0 wbl  (dof=5)
\item 2 exp + 1 wbl  (dof=6)
\end{itemize}

To quote Ref.\cite{Lubke2016} ``[t]he practice of using the same data set to select a best-fitting model and to assess the significance of model parameter estimates or interpret the model structure is based on the often implicit assumption that the selected model is the true model that generated the data [...]. However, this assumption does not hold in general. The sampling error related to model selection is ignored if the same data are used for inference.'' Thus, we separate the task of model selection from the best fit of the parameters.

\subsubsection{Model selection and bootstrapping}
For the model selection we take for every trading day in the months of June and July 2010  a random time stamp for each individual stock. From this time stamp onward we take 200 successive time stamps for the big stocks (RIO and BARC) and 100 successive time stamps for the small stocks (RRLN and ABFL). For each of these original samples we generate additional 999 bootstrap samples out of the original sample (e.g. for more details on bootstrapping the standard reference~\cite{Efron1979}) . For each ensemble of 1000 bootstrap samples we run the censored EM-algorithm and compute the log-likelihood and the BIC. Hence, for each model we have obtained a BIC distribution which is approximately normal. We then perform a Welch t-test with 5\% confidence level on the following hypothesis

{\bf ``Can the alternative model beat 1 exp + 1 Weibull using BIC?''}

We then count the success rate for the winning distribution. Our results are depicted in Table~\ref{tab:BIC_JuneJuly}, displaying the proportion of winnings. Our first intuition is confirmed:  the mixture ``1 exponential + 1 Weibull'' is the clear winner.
\begin{table}[htbp]
  \centering
  \caption{BIC-winners from bootstrapping ensembles in June and July 2010}
  \begin{ruledtabular}
    \begin{tabular}{lrrrrr}
    Model      & \multicolumn{1}{l}{dof} & \multicolumn{1}{l}{RIO} & \multicolumn{1}{l}{BARC} & \multicolumn{1}{l}{RRLN} & \multicolumn{1}{l}{ABFLN} \\\hline
    1 exp + 1 wbl & 4     & 0.77  & 0.77  & 0.77  & 0.76 \\
    0 exp + 2 wbl & 5     & 0.09  & 0.14  & 0.09  & 0.12 \\
    3 exp + 0 wbl & 5     & 0.14  & 0.09  & 0.14  & 0.12 \\
    2 exp + 1 wbl & 6     & 0     & 0     & 0     & 0 \\
    \end{tabular}%
    \end{ruledtabular}
  \label{tab:BIC_JuneJuly}%
\end{table}%


\subsubsection{Results for the preferred model:  ``1 exponential + 1 Weibull''}
In this subsection we use the convention that for the exponential contribution of the mixture $\beta_1=1$ will be suppressed and for the Weibull contribution we write $\beta$ rather than $\beta_2$ for easier reading.
In   Table~\ref{tab:beta} the results of the estimated parameters are summarised. We find for the ``complete'' data samples that the Weibull shape parameter $\beta$ takes on a universal value of approximately 0.57 as already found in Ref.\cite{Kizilersu2016} (where using a left-truncation was the way of handling the ``zero-inflated'' data).
\begin{table*}[htbp]
  \centering
  \caption{ Weibull component for ``1 exp + 1 Weibull''}
  \begin{ruledtabular}
     \begin{tabular}{lcccccc}
          & \multicolumn{1}{l}{$\frac{1}{N}\left(\log{L} \pm \Delta \log{L}\right)$} &  \multicolumn{1}{l}{weight} & \multicolumn{1}{l}{median $\beta$} & \multicolumn{1}{l}{ $\beta \pm \Delta \beta$} &  \multicolumn{1}{l}{ No. samples} \\ \hline
    RIO   & -8.29$\pm$0.63 & 0.82 & 0.55 & 0.57  $\pm$ 0.11 & 3107\\
    BARC  & -8.24$\pm$0.72 & 0.81 & 0.57 & 0.58  $\pm$ 0.11 & 3346\\
    RRLN  & -9.35$\pm$0.83 & 0.78 & 0.48  & 0.50  $\pm$ 0.12 & 535\\
   ABFLN & -9.91$\pm$0.78 & 0.81 & 0.47 & 0.49  $\pm$ 0.12 & 291 \\
    \end{tabular}%
    \end{ruledtabular}
  \label{tab:beta}%
\end{table*}%

\begin{table*}[htbp]
  \centering
  \caption{Scale parameters $\alpha$ in [ms] for  ``1 exp + 1 Weibull''}
  \begin{ruledtabular}
     \begin{tabular}{lcccccc}
          & \multicolumn{1}{l}{median $\alpha_{2}$} &  \multicolumn{1}{l}{range} & \multicolumn{1}{l}{median $\alpha_{1}$} & \multicolumn{1}{l}{range} &  \multicolumn{1}{l}{ No. samples} \\ \hline
    RIO   & 2499 & [1099,5491] & 17.2 & [6.5,70.8] & 3107\\
    BARC &2452  & [1078,5310] & 19.0 & [9.0,64.0] & 3346\\
    RRLN  &13846 & [6079,28988] & 16.0 & [6.6,47.1] &  535\\
   ABFLN & 23095 & [10374,45580] & 18.7 & [7.0,49.1] & 291 \\
    \end{tabular}%
    \end{ruledtabular}
  \label{tab:alpha}%
\end{table*}%

\begin{figure}[hptb]
	\begin{center}
		\includegraphics[width=0.5\textwidth]{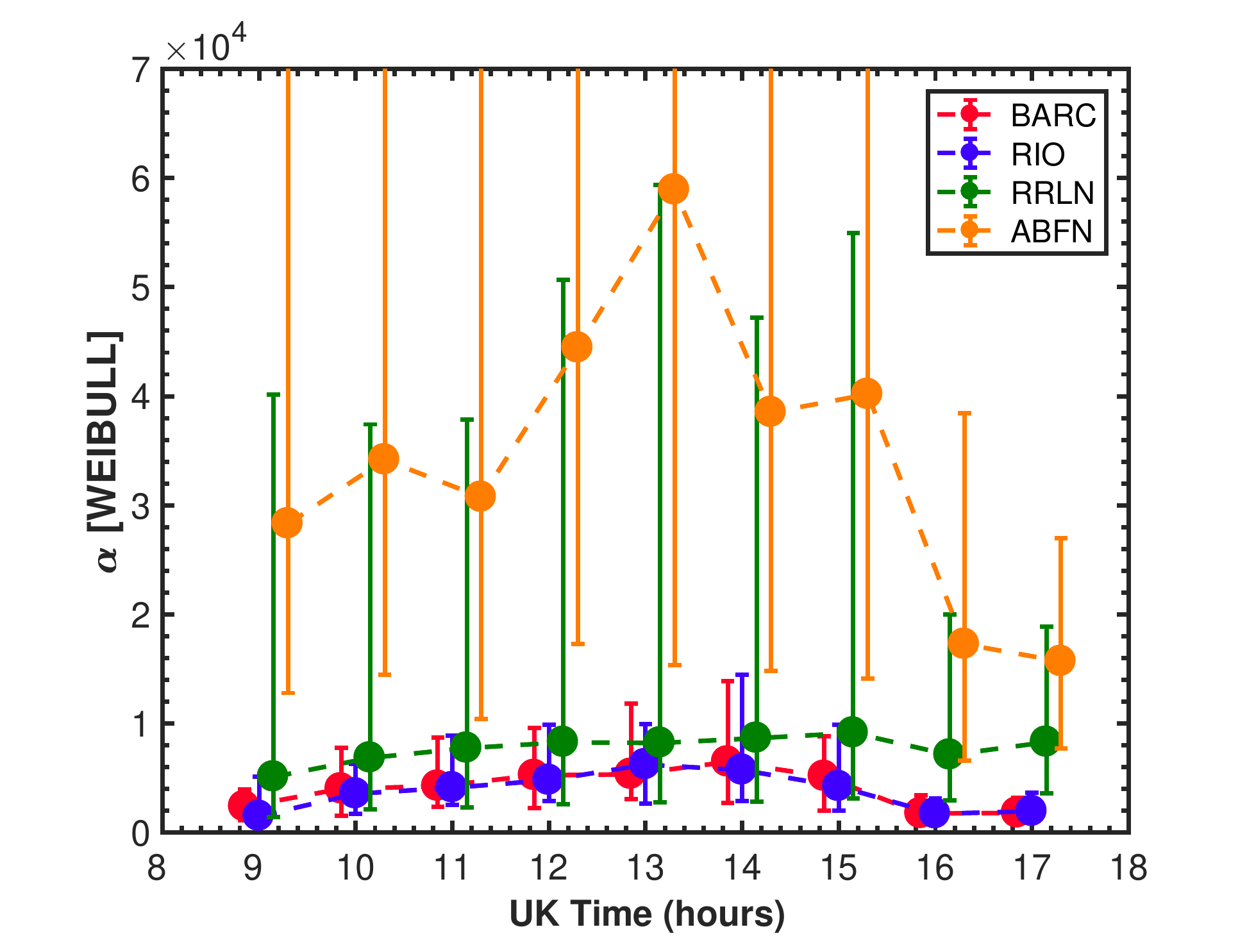}
		\caption{Weibull $\alpha_2$ in milliseconds for various tickers during trading hours.\label{fig:alphas}}
	\end{center}
\end{figure}

From Table~\ref{tab:alpha} we see that the median of the Weibull scale parameter $\alpha_2$ varies for different stocks (depending on their trading activity), whereas the median of the exponential scale parameter $\alpha_1$ seems to be the same for different stocks. From Table~\ref{tab:alpha} we suspect a stronger time dependence for the Weibull scale parameter. To investigate this time dependence, we divide the trading hours from 9:00 to 17:30 UK summer time in intervals of 10 (respectively 30) minutes for RIO and BARC (RRLN and ABFLN respectively) and average the values of $\alpha_2$, $\alpha_1$ and $\beta$ over the trading days from June to September 2010. Due to this partitioning of the data, the sample size will vary significantly, depending on the time of the day, with $N\gg 200$ at some times and $N\ll 200$ at some other times.
As we have already remarked in Ref.\cite{Kizilersu2016} the Weibull scale parameter $\alpha_2$ will exhibit a strong time dependence during the trading day. The more actively a stock is traded, the smaller the Weibull scale parameter $\alpha_2$ will be. We see from Figure~\ref{fig:alphas} that the typical scale parameter for the big stocks RIO and BARC is well below 10 seconds, and both curves as function of time are nearly identical. This can be explained by index arbitrage in the FTSE100, which requires trading in big stocks such as RIO and BARC at the same time to exploit the arbitrage. Obviously at lunch time there is less activity and the $\alpha_2$ becomes larger.

\begin{figure}[hptb]
	\begin{center}
		\includegraphics[width=0.5\textwidth]{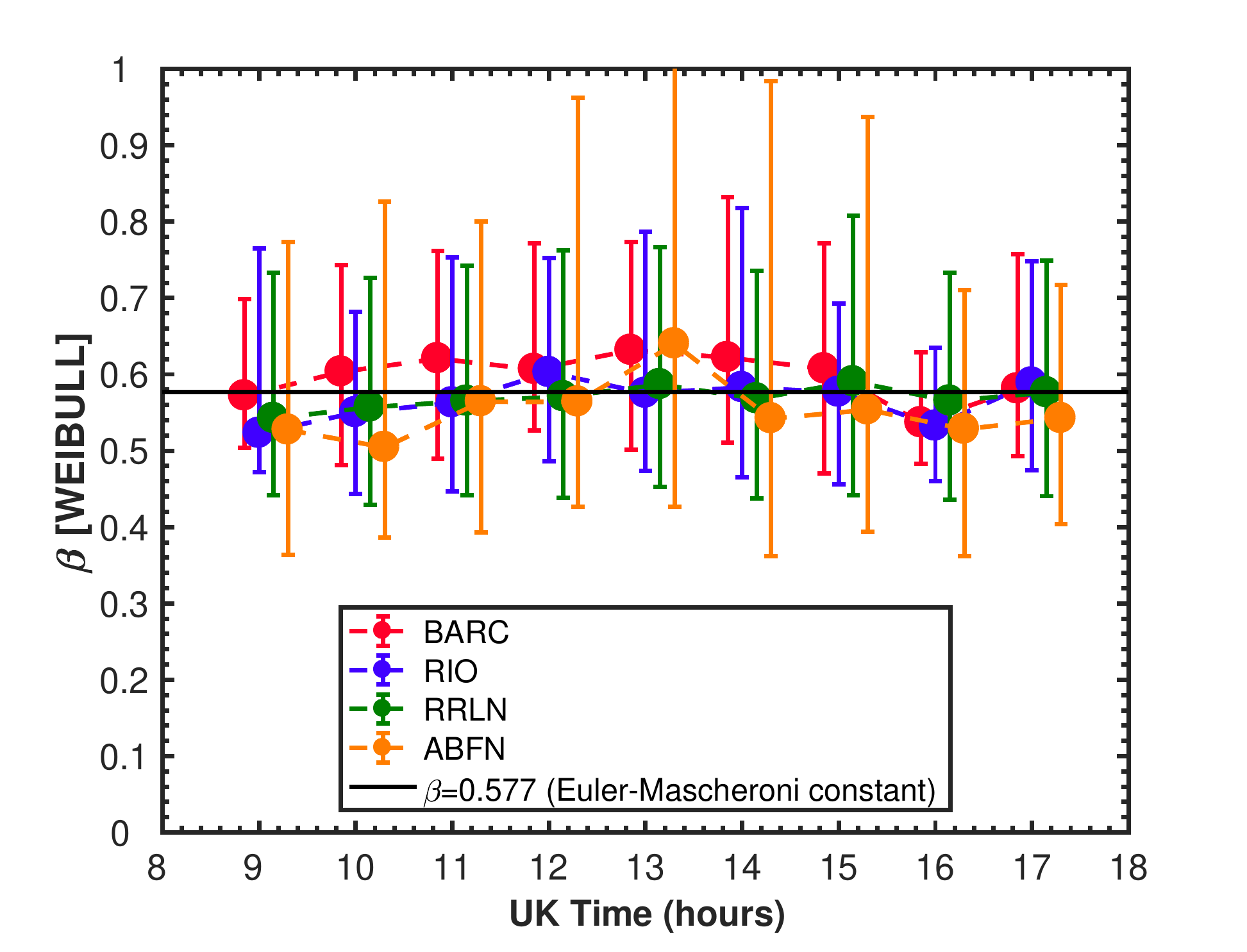}
		\caption{$\beta$ for various tickers during trading hours.\label{fig:betas}}
	\end{center}
\end{figure}
Since the Fisher information matrix is not diagonal for the MLE problem for Weibull distributions a bias in the estimated  scale parameter $\alpha_2$ will also result in a bias of the estimated shape parameter $\beta$. Thus, we see in Figure~\ref{fig:betas} that the value of $\beta$ becomes slightly larger at lunch time, when the estimated value of $\alpha_2$ is larger than that found at the opening or closing of trading hours. Despite this, the high level of stability found for $\beta$ suggests that it is reasonable to assume that the true shape parameter is universal with $\beta=0.57$ as conjectured in Ref.\cite{Kizilersu2016}.

\begin{figure}[hptb]
	\begin{center}
		\includegraphics[width=0.5\textwidth]{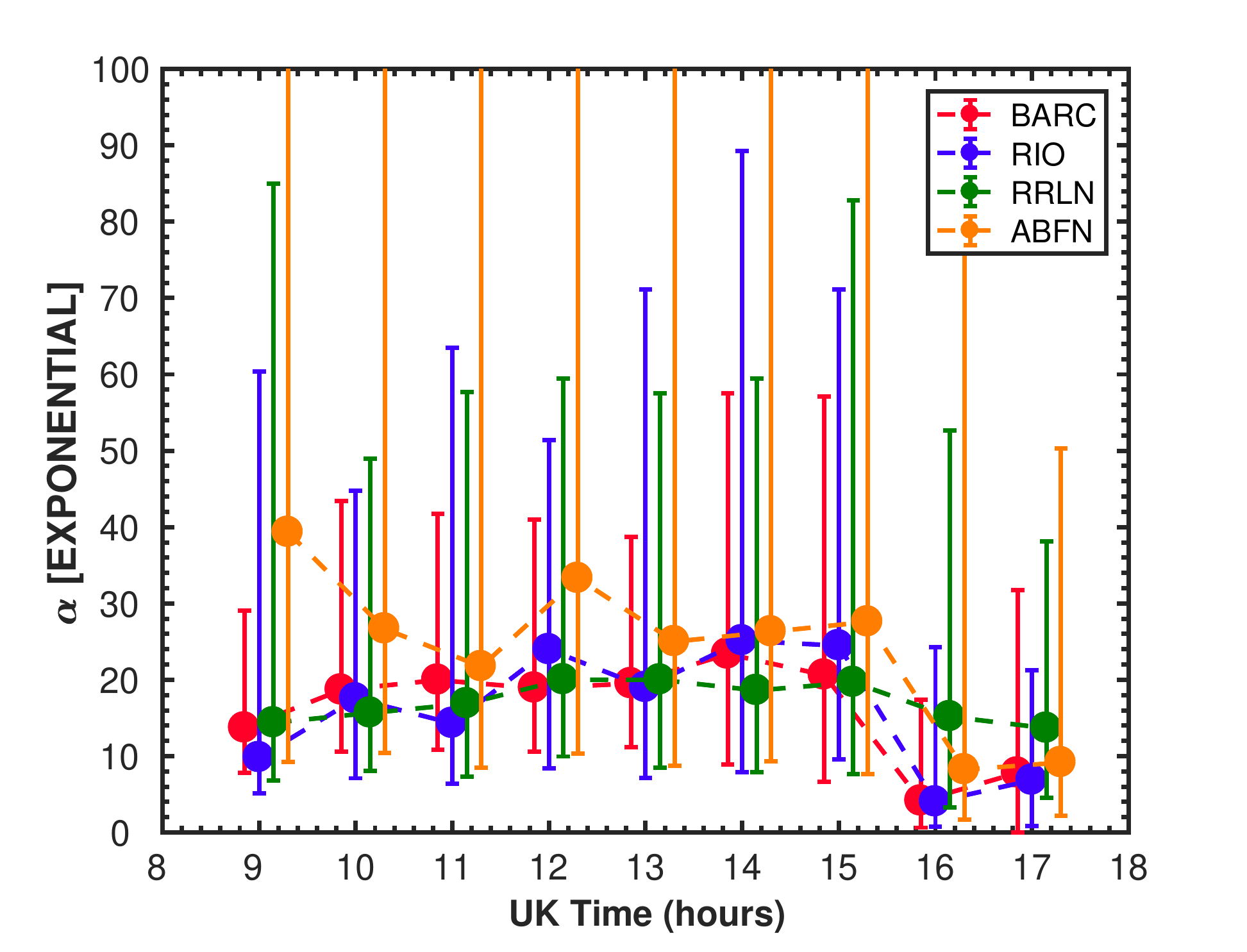}
		\caption{$\alpha_{1}$ in milliseconds for various tickers during trading hours.\label{fig:alphas_exp}}
	\end{center}
\end{figure}

\begin{figure}[hptb]
	\begin{center}
		\includegraphics[width=0.49\textwidth]{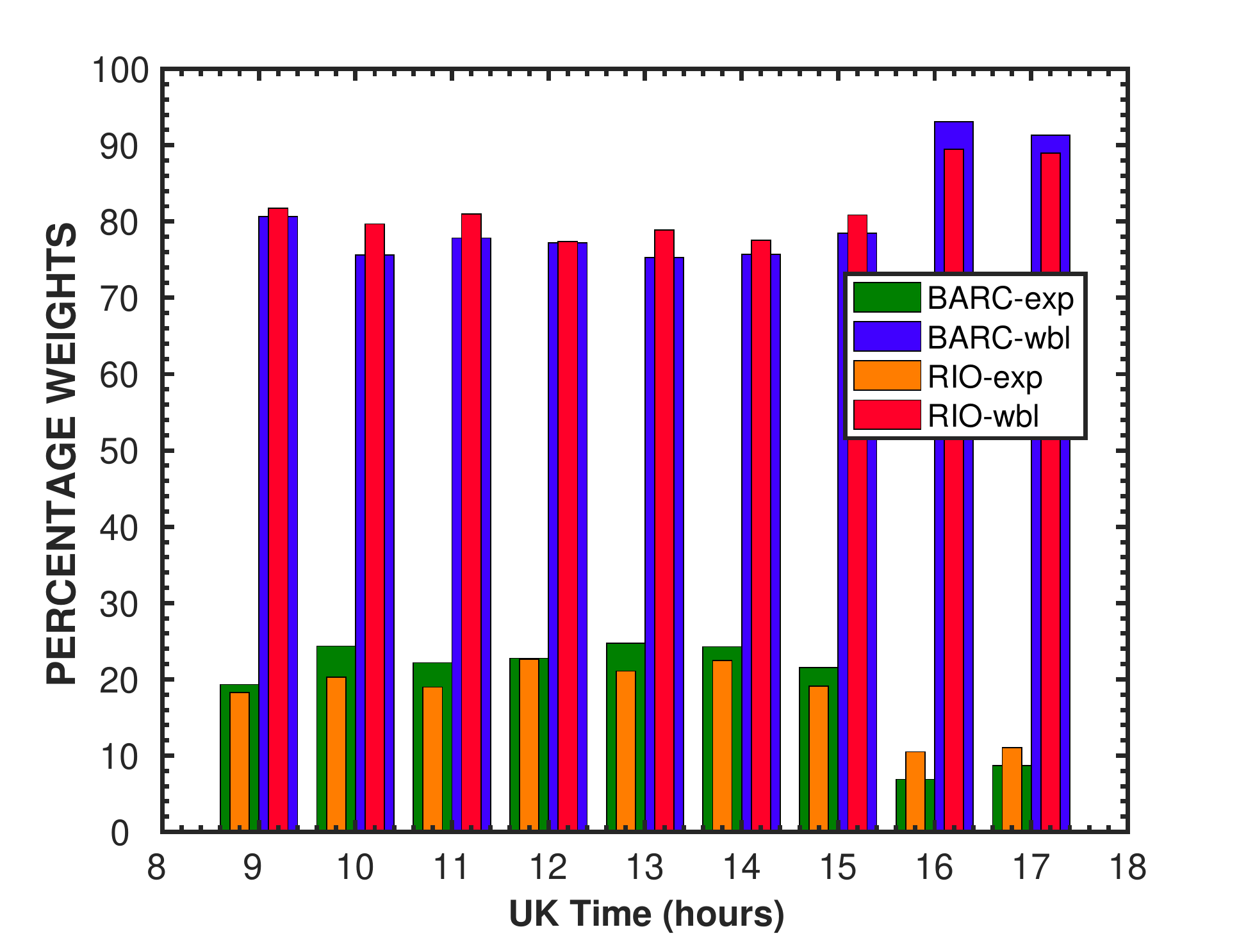}
		\includegraphics[width=0.49\textwidth]{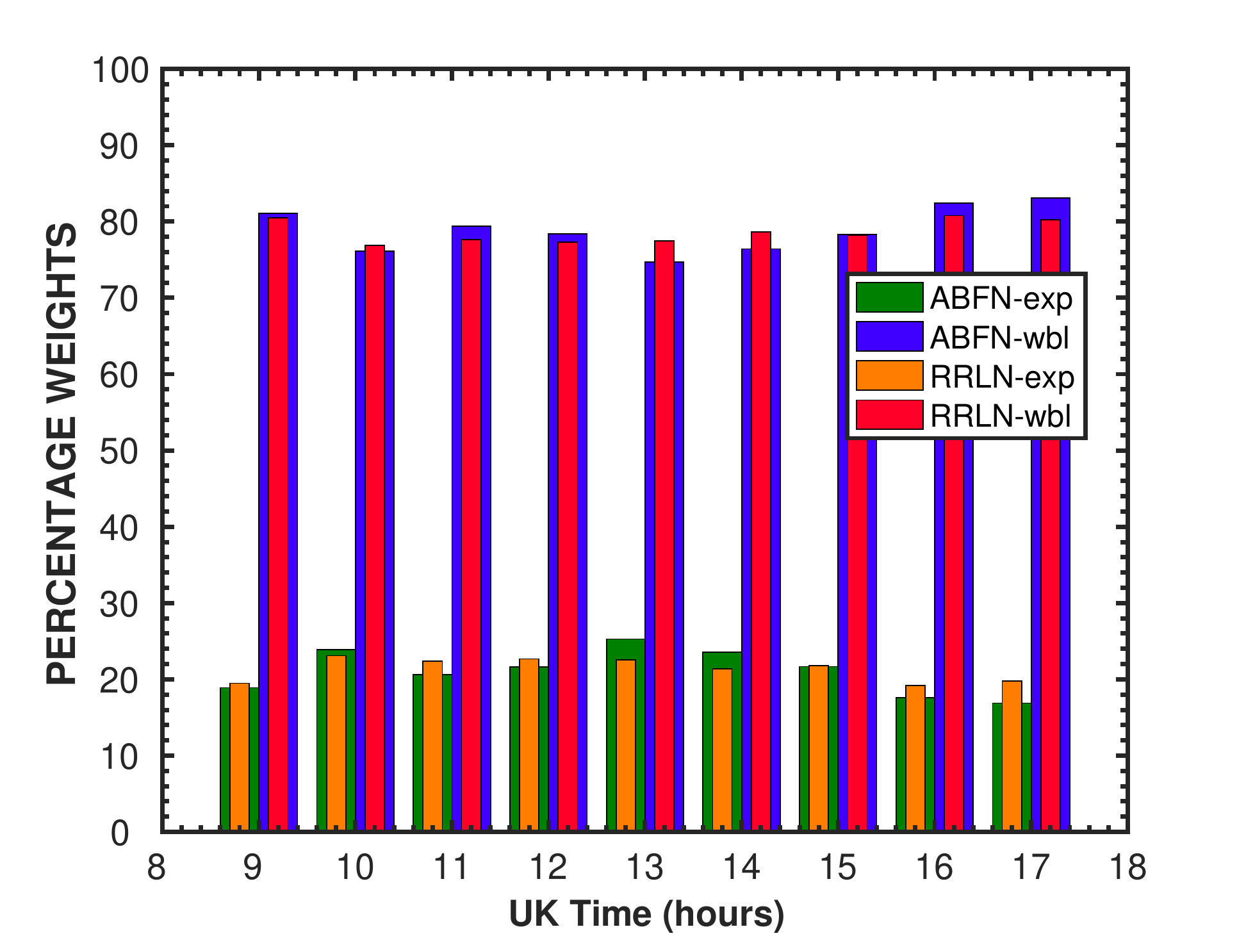}
		\caption{Mixture weights for various tickers during trading hours.\label{fig:weights}}
	\end{center}
\end{figure}

Finally we point the reader's attention to Fig.~\ref{fig:alphas_exp} which displaying the time dependence of the scale parameters $\alpha_1$ of the exponential contribution and Fig.~\ref{fig:weights} displaying the time dependence of the mixture weights: the time dependence of $\alpha_1$ is mainly in the region of 10 to 20 milliseconds for all stocks with exceptions of the ``smallest'' and sometimes illiquid stock ABFLN. Also the weights of this exponential contribution in the two-component mixture is nearly independent of time and consistently around 20 \%, valid for all stocks.

\section{Conclusion}
The censored EM-algorithm in combination with a bootstrapping argument applied to the Baysean Information Criterion (BIC) allows us to choose as a model for  the MO arrival times a two-component mixture distribution consisting of ``1 exponential + 1 Weibull''. Of course, this conclusion is not applicable in the exceptional case  of extraordinary trading activity in a stock when critical information is disclosed to the market participants. 
The first component of this mixed distribution is an exponential distribution with a relative weight of approximately 20\% and a rather short scale parameter, in the range of 10 to 20 milliseconds. This result is independent of the stock under consideration and almost constant during the trading day. The second component, with a relative weight of approximately 80\%, is a Weibull distribution for which the scale parameter varies intra-day with trading activity and lies typically between 1000 and 25000 milliseconds, albeit with universal shape parameter $\beta \approx 0.57$. 

This result can be understood with a view to a typical stock exchange computer architecture (e.g. Ref.\cite{Lovelace2013}): market orders are captured via various order gateways which forward the electronic orders to the so called accumulator,  where timestamps are given. Low latency order gateways typically cater for high-frequency traders (e.g. algorithmic hedge funds) whose buy- and sell-orders are generated by computers located in the stock exchange's immediate neighbourhood to minimize the transmission time. Here an exponential waiting time between these signals (with a short time scale parameter) is to be expected resulting in a Poisson process for high-frequency buy- and sell-orders. On the other hand, VSAT gateways cater for stock brockers whose customers include both institutional and private clients and whose orders are usually transmitted via an internet connection. It has been shown that internet traffic sends information as TCP-``parcels'' with arrival times that can be described by Weibull distributions with a shape parameter less than 1 (see Ref.\cite{Feldmann2002}, Ref.\cite{Arfeen2019}). In Ref.\cite{Feldmann2002} it was shown that a shape parameter $\beta=0.569$ provided an excellent fit to the observed data. This is close to the value that we find for the Weibull distribution. The natural conclusion would be that for the time period we have studied trading on the LSE approximately 10\% of the trading was done by market participants whose orders were generated electronically by a computer in the immediate neighborhood, whereas the other 90\% of market participants were using the internet to generate their buy- and sell-orders.  

Finally we want to emphasize that by Ref.\cite{Yannaros1994} any Weibull distribution function with scale parameter $\beta$ smaller than 1 can be expressed as a superposition of exponential waiting time distributions. Thus our favorite model ``1 Weibull + 1 exponential'' is equivalent to a suitable mixture of exponential waiting time distributions. Consequently, the proposed censored EM algorithm for finite Weibull mixtures maybe compared to the problem discussed firstly in Ref.\cite{Scalas2007} and later expanded in Ref.\cite{Ponta2019} of how to fit a waiting time distribution consisting of a finite number of exponential waiting time distributions to the observed tick-by-tick data. 
Although exponential mixtures with more than 5 components seem to describe our actual data sets very well, they are excluded by the BIC due to their too high dof-value. 

\appendix 
\section{Expression of objective function in M-step and further auxilary functions}

First define the censoring intervals $\mathcal{I}_{\ell}=[\xi_{\ell-1},\xi_{\ell})$. For simplicity introduce a transformed quantity $\zeta_{\ell} = (\xi_{\ell}/\alpha_i^{(k-1)})^{\beta_i^{(k-1)}}$. Then we find after some lengthy computations in the spirit of section 2  the following version of Eq.(\ref{eq:Q_term})

\begin{widetext}
\begin{eqnarray}
&& \sum_{j=1}^{n} z_{1j}^{(k)} \left[ \log{\frac{1}{\alpha_1^{(k)}}}-\frac{x_j}{\alpha_1^{(k)}}\right] +  \sum_{\ell=1}^L N_{\ell}~ \tilde{z}_{1\ell}^{(k)} \Bigg[ \log{\frac{1}{\alpha_1^{(k)}}}-\frac{1}{\alpha_1^{(k)}}C_{1\ell}^{(k-1)} \Bigg]
\nonumber 
\\
&+&
\sum_{j=1}^{n} z_{2j}^{(k)} \left[ \log{\frac{\beta_2^{(k)}}{\alpha_2^{(k)}}}+(\beta_2^{(k)}-1)\log{\frac{x_j}{\alpha_2^{(k)}}}  - \left(\frac{x_j}{\alpha_2^{(k)}}\right)^{\beta_2^{(k)}} \right]
\nonumber
\\
&+&  \sum_{\ell=1}^L N_{\ell}~ \tilde{z}_{2\ell}^{(k)} \Bigg\{
 \log{ \frac{\beta_2^{(k)}}{\alpha_2^{(k)} }} + (\beta_2^{(k)}-1) \log{ \frac{\alpha_2^{(k-1)}}{\alpha_2^{(k)} }} 
\nonumber
\\
&+ & \left. \frac{\beta_2^{(k)}-1}{\beta_2^{(k-1)} } ~ \frac{e^{-\zeta_{\ell-1}}\log{\zeta_{\ell-1}} - e^{-\zeta_{\ell}}\log{\zeta_{\ell}}+ \Gamma( 0, \zeta_{\ell-1}) -\Gamma(0,\zeta_{\ell})}{e^{-\zeta_{\ell-1}}-e^{-\zeta_{\ell}}}  \right.
\nonumber
\\
&{\color{red} - }&  \left(\frac{\alpha_2^{(k-1)}}{ \alpha_2^{(k)}} \right)^{\beta_2^{(k)}} \frac{ \Gamma\left(\frac{\beta_2^{(k)}}{\beta_2^{(k-1)}}+1 ,\zeta_{\ell-1}\right)-\Gamma\left(\frac{\beta_2^{(k)}}{\beta_2^{(k-1)}}+1 ,\zeta_{\ell}\right) }{e^{-\zeta_{\ell-1}}-e^{-\zeta_{\ell}}}
 \Bigg\}
\nonumber
\\
\label{eq:MLE}
\end{eqnarray}
\end{widetext}
where
\begin{eqnarray}
C_{1\ell}^{(k-1)} = \alpha_1^{(k-1)} + \frac{\xi_{\ell-1}e^{-\frac{\xi_{\ell-1}}{\alpha_1^{(k-1)}}} -\xi_{\ell}e^{-\frac{\xi_{\ell}}{\alpha_1^{(k-1)}}}}{e^{-\frac{\xi_{\ell-1}}{\alpha_1^{(k-1)}}}-e^{-\frac{\xi_{\ell}}{\alpha_1^{(k-1)}}}}
\label{eq:C_ell}
\end{eqnarray}
Note that in Eq.(\ref{eq:MLE}) we have for the lower censoring interval boundary $\zeta_0=0$  that the term with $\ell=1$ is well-behaved and reduces to
\begin{eqnarray}
&&\frac{e^{-\zeta_{\ell-1}}\log{\zeta_{\ell-1}} - e^{-\zeta_{\ell}}\log{\zeta_{\ell}}+ \Gamma( 0, \zeta_{\ell-1}) -\Gamma(0,\zeta_{\ell})}{e^{-\zeta_{\ell-1}}-e^{-\zeta_{\ell}}} \nonumber\\
&& \vspace*{5cm} = -\frac{\gamma + e^{-\zeta_{1}}\log{\zeta_{1}}+\Gamma(0,\zeta_{1})}{1-e^{-\zeta_{1}}}
\nonumber
\end{eqnarray}

We need to maximise expression Eq.(\ref{eq:MLE}) with respect to the parameters $\alpha_1^{(k)},\alpha_2^{(k)},\beta_2^{(k)}$, all other quantities being known from previous steps. 

For the MLE equations  the following expression arises during the lengthy computations
\begin{widetext}
\begin{eqnarray}
D_{2\ell}^{(k)}& =& 
\frac{e^{-\zeta_{\ell-1}}\log{\zeta_{\ell-1}}- e^{-\zeta_{\ell}}\log{\zeta_{\ell}}  + \Gamma(0,\zeta_{\ell-1})-\Gamma(0,\zeta_{\ell})}{\beta_2^{(k-1)}}
\nonumber
\\
&&- \frac{1}{\beta_2^{(k-1)}} \left( \frac{\alpha_2^{(k-1)}}{\alpha_2^{(k)}} \right)^{\beta_2^{(k)}}
\Bigg[  \sum_{p=0}^{\infty} \frac{(-1)^p}{p!} \frac{\zeta_{\ell-1}^{\frac{\beta_2^{(k)}}{\beta_2^{(k-1)}}+1+p}-\zeta_{\ell}^{\frac{\beta_2^{(k)}}{\beta_2^{(k-1)}}+1+p}}{\left(\frac{\beta_2^{(k)}}{\beta_2^{(k-1)}}+1+p\right)^2} 
\nonumber
\\
& &  - \Gamma \left(\frac{\beta_2^{(k)}}{\beta_2^{(k-1)}}+1 \right) \left(\log{\zeta_{\ell-1}}-\log{\zeta_{\ell}}\right)
\nonumber
\\
&& +\Gamma \left(\frac{\beta_2^{(k)}}{\beta_2^{(k-1)}}+1 ,\zeta_{\ell-1}\right) \log{\zeta_{\ell-1}} - \Gamma \left(\frac{\beta_2^{(k)}}{\beta_2^{(k-1)}}+1 ,\zeta_{\ell}\right) \log{\zeta_{\ell}} \Bigg]
\nonumber
\\
& & - \log{\frac{\alpha_2^{(k-1)}}{\alpha_2^{(k)}}} \left(\frac{\alpha_2^{(k-1)}}{\alpha_2^{(k)}}\right)^{\beta_2^{(k)}}
\Bigg\{ \Gamma \left(\frac{\beta_2^{(k)}}{\beta_2^{(k-1)}}+1 ,\zeta_{\ell-1}\right) - \Gamma \left(\frac{\beta_2^{(k)}}{\beta_2^{(k-1)}}+1 ,\zeta_{\ell}\right) \Bigg\}
\nonumber
\\
\label{eq:Weibull_aux}
\end{eqnarray}
\end{widetext}
and for $\ell=1$
\begin{widetext}
\begin{eqnarray}
D_{21}^{(k)}& =& 
\frac{-\gamma- e^{-\zeta_{1}}\log{\zeta_{1}}  -\Gamma(0,\zeta_{1})}{\beta_2^{(k-1)}}
\nonumber
\\
&&- \frac{1}{\beta_2^{(k-1)}} \left( \frac{\alpha_2^{(k-1)}}{\alpha_2^{(k)}} \right)^{\beta_2^{(k)}}
\Bigg[ - \sum_{p=0}^{\infty} \frac{(-1)^p}{p!} \frac{\zeta_{1}^{\frac{\beta_2^{(k)}}{\beta_2^{(k-1)}}+1+p}}{\left(\frac{\beta_2^{(k)}}{\beta_2^{(k-1)}}+1+p\right)^2} 
\nonumber
\\
& &  + \Gamma \left(\frac{\beta_2^{(k)}}{\beta_2^{(k-1)}}+1 \right) \log{\zeta_{1}}
  - \Gamma \left(\frac{\beta_2^{(k)}}{\beta_2^{(k-1)}}+1 ,\zeta_{1}\right) \log{\zeta_{1}} \Bigg]
\nonumber
\\
& & - \log{\frac{\alpha_2^{(k-1)}}{\alpha_2^{(k)}}} \left(\frac{\alpha_2^{(k-1)}}{\alpha_2^{(k)}}\right)^{\beta_2^{(k)}}
\Bigg\{ \Gamma \left(\frac{\beta_2^{(k)}}{\beta_2^{(k-1)}}+1 \right) - \Gamma \left(\frac{\beta_2^{(k)}}{\beta_2^{(k-1)}}+1 ,\zeta_{1}\right)  \Bigg\}
\nonumber
\\
\label{eq:Weibull_aux0}
\end{eqnarray}
\end{widetext}

\bibliography{finance}
\end{document}